\documentclass[onecolumn]{aa}
\usepackage{graphicx}
\usepackage{txfonts}
\begin{document}
   \title{First detection of a lithium rich carbon star in the Draco dwarf galaxy:
   evidence for a young stellar population}

   \subtitle{}

   \author{I. Dom\'\i nguez\inst{1}, C. Abia\inst{1}, O. Straniero\inst{2}, S. Cristallo\inst{2}
          \and
          Ya.V. Pavlenko\inst{3}
          }

\offprints{I. Dom\'\i nguez, inma@ugr.es}

\institute{Dpto. F\'\i sica Te\'orica y del Cosmos,
              Universidad de Granada, 18071 Granada, Spain
              \and
              INAF-Osservatorio Astronomico di Collurania, I-64100 Teramo, Italy
             \and              
             Main Astronomical Observatory, National Academy of Sciences,
              Zabolotnoho 27, Kiev-127 03680, Ukranie
             }

   \date{Received .......; accepted .........}

   \abstract{We present a spectroscopic study of D461, a giant star belonging to Draco
dwarf spheroidal galaxy. 
 From spectral synthesis in LTE
we derive a lithium  abundance of $\rm log$ $\epsilon$(Li)$=3.5\pm0.4$ and a C/O ratio between 3 and 5. 
This is 
the first detection of a lithium rich C-star in a dwarf spheroidal galaxy. 
Basing on stellar models of appropriate chemical composition,
we show that a similar C enrichment is compatible with that expected
for a low mass low metallicity thermally pulsing AGB star, undergoing few third dredge up episodes.
The position in the $\rm log$ g-$\rm log$ T$_{\rm{eff}}$ diagram of  D461 is also compatible with this 
theoretical scenario. In particular, the low effective temperature, lower
 than that expected for a low metallicity giant star, is a consequence of the huge
increase of the envelope opacity occurring after the carbon dredge up.     
The Li enrichment may be explained if
a deep circulation would take place during the interpulse period, the so called cool
 bottom process. In spite of the low resolution of our spectra,  
we derive a lower limit for the carbon isotopic ratio, 
namely $^{12}$C/$^{13}$C$>40$, and a constraint for the Ba abundance, namely
$0.5<$[Ba/Fe]$< 2$. The proposed scenario also fits these further constraints.
Then, we estimate that the mass of D461 ranges between 1.2 and 2 $M_\odot$, which 
corresponds to 
an age ranging between 1 and 3 Gyr. We conclude that 
this star is more massive and younger than the typical stellar population of Draco.
   
}
   
\titlerunning{First detection of a lithium rich carbon star in Draco}
\authorrunning{I. Dom\'\i nguez et {\em al.}}

\maketitle
%

\section{Introduction}
 The origin of the Galactic lithium is still not completely understood. In addition to that
produced by the primordial nucleosynthesis, several sources 
have been invoked to explain the observed abundances in the various components of the Milky Way 
(see Travaglio et al. 2001, for a recent analysis of the contributions of various
 possible producers of galactic Li).
Among these, only for asymptotic giant branch (AGB) stars there are
observational evidences of some Li production.
 Li is easily destroyed within stars due to the large
cross section of its proton capture reaction. As a consequence, giant stars become 
Li-depleted after the first dredge up.
However, it has been early recognized that a Li production may occur in AGB stars
via the beryllium convective belt mechanism (Cameron \& Fowler 1971). Two are the conditions to be
fulfilled: i) the temperature at the base of the convective envelope must be of the order of 
20-30 $10^6$ K, so that $^7$Be can be produced via $^3$He$+^4$He reaction, and ii) the mixing
must be fast enough to remove the fresh Be from the hot bottom layers,
before it decays into Li by electron capture. In such a way, most
of the Li will be synthesized in the cool external layers of the star, where the proton 
capture are defused. 
Several AGB stars with large abundances of Li 
 are actually found in the Galaxy and 
 in the Magellanic Clouds
(Catchpole \& Feast 1976; Abia et al. 1993; Smith et al. 1995).

Li enhancement is commonly found in stellar models of
massive AGB stars as a consequence of the hot bottom burning 
(HBB; Renzini \& Voli 1981; Sackmann \&
Boothroyd 1992; Forestini \& Charbonnel 1997;  Lattanzio \& Forestini 1999). However,
there are clear indications of an efficient Li production in some low mass AGB stars,
where the HBB does not occur. 
For instance, most of the SLiR \footnote{We define lithium rich (LiR) and super lithium rich
(SLiR) those stars showing log $\epsilon$(Li)$>1.0$ and log $\epsilon$(Li)$>4.0$, respectively;
as usual, log $\epsilon$(X) = log (X/H)+12, where X/H  is the number of atoms
 of a given element relative to hydrogen.} AGB stars found in the
Galaxy are carbon stars (C/O$>1$ in the envelope, see however Garc\'\i a-Lario et al. 1999), 
while, in the Magellanic Clouds, they are mostly O-rich. Moreover, the luminosities
of LiR and SLiR stars in the Clouds are systematically brighter ($-6\leq \rm{M_{bol}}\leq -7$) 
than their galactic counterparts (M$_{\rm{bol}}\geq -5.5$). Their low luminosity and 
 the fact that C/O$>1$ indicate that most of the galactic Li rich stars have low mass progenitors. 
Finally, the relative abundance of galactic LiR stars  with low $^{12}$C/$^{13}$C ratios
($< 15$, mostly C-stars of type J; Abia \& Isern 1997) 
is substantially larger than that measured in the Magellanic Clouds    
(Hatzidimitriou et al. 2003).
In summary, the comprehension of the mechanisms of Li production in different 
AGB stars as well as
its dependence on the parent stellar population deserve further observational and theoretical
 investigations. 

Dwarf galaxies are other stellar systems where AGB stars can be resolved.
 In the last few years,
a growing amount of AGB stars have been discovered in these galaxies: 
Withelock et al. (1999) in Sagittarius; Azzopardi et al. (1999) and Demers et al. (2002) in
Fornax; Shetrone et al. (2001a) and Margon et al. (2002) in Draco (see also Groenewegen 1999 and
references therein). Dwarf galaxies span
a wide range in metallicity, so that they could provide new hints about the dependence of the
AGB nucleosynthesis (Li production included) on the chemical composition of the parent
stellar population.

As part of a large survey of AGB stars in the nearby galaxies, we report here,
the spectroscopic study of the carbon star D461 in the Draco dwarf spheroidal galaxy, 
discovered by Armandroff et al. (1995). Let us anticipate our main findings:
a) D461 is Li-rich and b)it is more massive and younger than the dominant stellar population of
Draco.   
This is the first report of a 
star of this type in a dwarf spheroidal.
In section 2, we present the observations and the abundance analysis.
The evolutionary state of D461 is investigated in section 3,  
by comparing the available photometric
and spectroscopic data to appropriate 
models of AGB stars. Possible evolutionary scenarios 
are discussed in the conclusive section.
 
\section{Observations, lithium identification and analysis}

Intermediate resolution spectra ($R\sim 6500$) of several C-stars in Draco were
obtained with the
ISIS spectrograph attached to the 4.2 m William Herschel telescope at the Roque
de los Muchachos
Observatory. The observations were made during July 4-5 2003. 
The spectral range covered was $6450-8100$ {\AA}. D461 was observed during
$\sim 4600$ s in
several exposures to minimize the impact of the sky background and cosmic rays.
The reduction and analysis
of the individual spectra was made with the standard techniques using IRAF. The
final spectrum has a S/N$\sim 60$ in the
region of the Li I $\lambda 6708$ {\AA} line.  

Few notices about Draco 461 (17$^h$ 19$'$ 42.40$''$;
+57$^\circ$ 58$'$ 37.8$^{''}$, Eq. 2000) can be found in the current literature. Our low
resolution spectra confirm previous identifications of the carbon-rich nature of this star
(Armandroft et al. 1995). Very strong CN
bands are clearly seen red-ward to 7850 {\AA} as well as
weaker CN bands in the  7000-7200 {\AA} region.  
Assuming, for D461, the visual magnitude given by Shetrone et al. (2001a), V=17.19, a distance modulus
  to Draco of $(m-M)_o=19.84\pm 0.14$ (Bellazzini et
al. 2002) and an interstellar extinction $E(B-V)=0.03$
(Mateo 1998), we derive M$_{\rm{V}}=-2.74\pm0.14$ for D461. Shetrone et al.
(2001a) give $(B-V)=1.74$ for this star, which
is the reddest color index among the carbon stars so far identified in the
Draco galaxy. However, D461 is significantly bluer
than the majority of the normal (N-type) carbon stars (therafter, C(N) stars) in the Galaxy.
The low metallicity of the Draco stellar population 
certainly plays a role in determining the photometric properties of its C-stars.
Several authors agree for an average metallicity $<$[Fe/H]$>=-2.0$ 
(Carney \& Seitzer 1986; Bell 1985;
Aparicio et al. 2001; Bellazzini et al. 2002), although there is a large
dispersion, $-3.0<$[Fe/H]$<-1.5$ (Shetrone et al. 1998).
The lack of an IR photometry also limits the knowledge of the effective temperature of D461.

We have evaluated the radial velocity of D461 by cross-correlating the observed
spectra with a theoretical spectrum. This has been done by using 
the $\lambda 6497$ {\AA} Ba II line, the
$\lambda 6563$ {\AA} H$_\alpha$ line (which is in {\it absorption}), 
and the CN band heads at $\lambda 7853$, $\lambda 7876$, $\lambda 7898$ and
$\lambda 7915$ {\AA}, respectively. The resulting mean radial velocity,  
 $V_r=-301.2 \pm 3.0$ kms$^{-1}$, is in good agreement with the value
obtained by Armandroft et al. (1995), $V_r=-299.9 \pm 1.5$ kms$^{-1}$. 
After correcting the observed spectra for the radial velocity, the strong
feature in the lithium region is placed at $\lambda 6707.804$ {\AA}. 
Considering the uncertainty in the wavelength calibration ($\pm 10$ m{\AA}), we
unambiguously identify it as the Li I doublet. There are little chances to 
confuse this feature with the Ce II line at $\lambda 6708.099$ {\AA} 
(Reyners et al. 2002). In this case, in fact, the equivalent width we derive for this line 
($\sim 1$ {\AA}) would imply an unrealistically large Ce abundance and other 
very strong features of s-process elements would emerge in the
spectrum.  

   \begin{figure*}
   \centering
   \includegraphics[angle=-90, width=17cm]{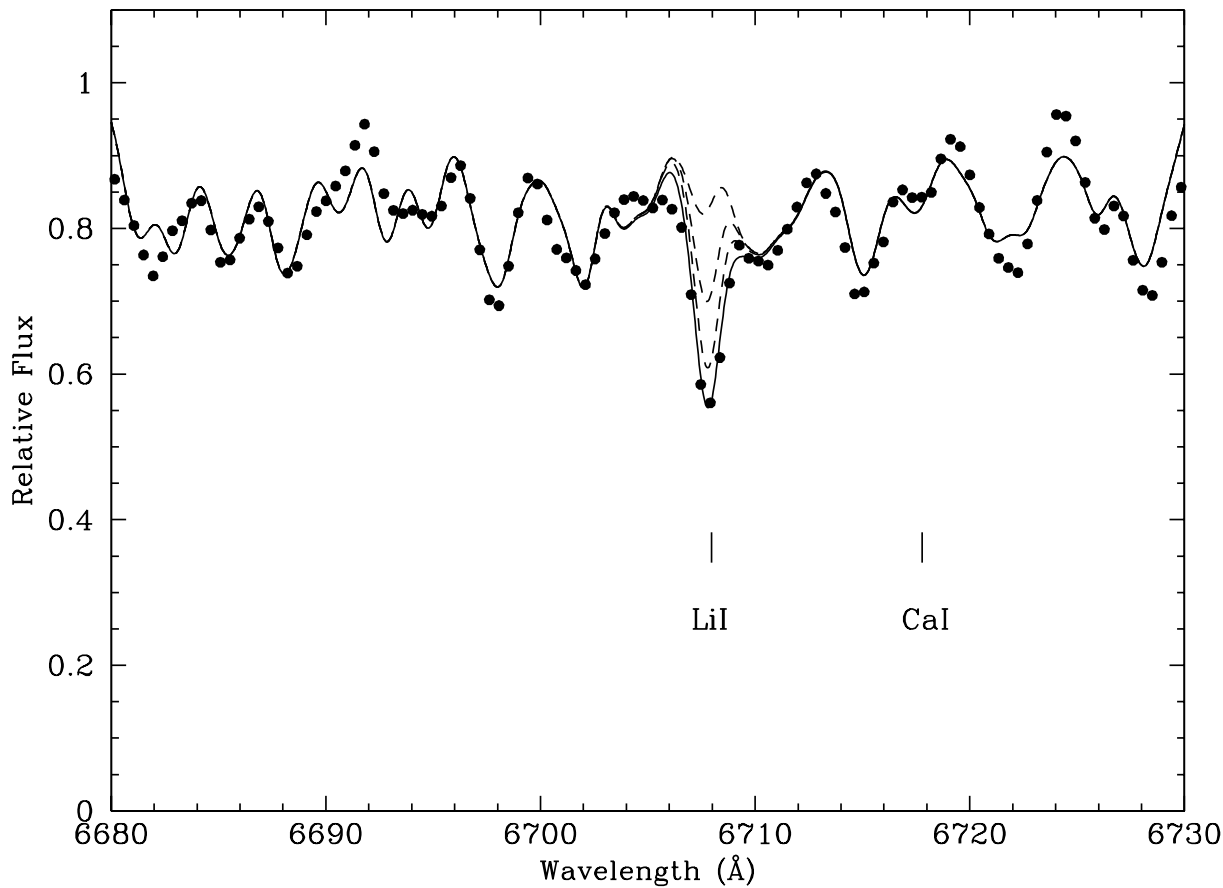}
   \caption{Synthetic fit to the spectrum of D461 in the lithium region. Solid
dots represent the observed spectrum.
Lines are synthetic spectra for different Li abundances: no Li, log
$\epsilon$(Li)=$1.5$ and 3.0 (dashed lines) and 3.5 (best fit,
continuous line). Theoretical spectra are computed assuming
T$_{\rm{eff}}=3600$ K, [Fe/H] $=-2.0$ and C/O$\sim 5$. Note that
the CaI line at $\sim\lambda 6718$ {\AA} is barely seen which it is an evidence
of the low metallicity of D461.}
\label{}%
\end{figure*}

The chemical analysis of D461 is complicated by the lack of a firm determination of its
effective temperature. In addition, the low resolution of our spectra does not allow the detection 
 of any unblended metal line to derive the metallicity. 
Because these difficulties, we computed a grid of model 
atmospheres for C-rich giants by means of the SAM12 program (see Pavlenko 2003 for
details of this code).  
To
account for the frequency and depth distribution of opacity in the
atmospheres of late-type stars we used an opacity sampling
approach (Sneden et al 1976). A few line lists from different
sources were taken into account:
 i) Line data of
diatomic molecules (CN, C$_2$, MgH, H$_2$, SiH, CO, CH, NH)
were taken from CDROM 18 of Kurucz (1993); ii) Atomic lines data were taken from VALD (Kupka et al. 1999). 
iii) HCN and HNC line lists were computed by
Harris et al. (2002); iv) Absorption of bands systems of CaO (C$^1\Sigma$ -
X$^1\Sigma$),  CS(A$^1\Sigma$ - X$^1\Sigma$), SO (A$^3\Pi$ -
X$^3\Sigma$), SiO (E$^1\Sigma$-X$^1\Sigma$), SiO(A$^1\Pi$ -
X$^1\Sigma^+$, NO (C$_2 \Pi_r$- X$_2\Pi_r$), NO(B$_2\Pi_r$
-X$_2\Pi_r$), NO(A$^2\Sigma^+$ - X$_2\Pi_r$), MgO(B$^1\Sigma^+$ -
X$^1\Sigma^+$), AlO(C$^2\Pi$ -X$^2\Sigma$), AlO(B$^2\Sigma^+$ -
X$^2\Sigma^+$) are taken into account by the JOLA approach (Nersisyan et al. 1986).

The following values of the model parameters have been adopted:
effective temperature between 3000
and 4000 K (with a step of 200 K), C/O$=$1.05, 1.10, 3, 5, and 7, metallicity
[Fe/H]= $-1.5, -1.7$ and $-2.0$, log $g=0$ and microturbulence $\xi=2.5$ kms$^{-1}$.
Then, we have tried to reproduce the spectrum of D461 in the
observed spectral range.
As a general rule,
the lower the effective temperature, the lower the
C/O ratio needed to fit the spectrum. Similarly, the higher the
metallicity of the model atmosphere, the lower the required C/O ratio. 
After an iteration process, the best reproduction of the observed
spectrum of D461 has been obtained for the following values of the model
atmosphere parameters: T$_{\rm {eff}}=3600\pm200$ K, [Fe/H] $=-2.0\pm0.2$ and
C/O between 3 and 5. 
The estimated value of T$_{\rm{eff}}$ is consistent with previous determinations
of the effective temperatures for other C-stars in Draco.  From infrared photometry,
Aaronson \& Mould (1985) derived 
T$_{\rm{eff}}=3800$, 3900 and 4100 K for the stars named C1, C2 and C3, respectively.
These C-stars of Draco have bluer $(B-V)$ index
than D461, which is consistent with the lower T$_{\rm{eff}}$ 
value we estimate for it. Note that the clear detection 
of the H$_\alpha$ line in absorption allow us to put a lower limit to
the effective temperature. The
cross-correlation of the observed spectra with the theoretical one, ruled out
the 
possibility that this absorption could have a circumstellar origin. 
Therefore, assuming a photospheric origin for this spectral feature, 
our tests have revealed that only for
T$_{\rm{eff}}\geq 3400$ K, the H$_\alpha$ line
 appears in absorption with an appreciable intensity.
We recall that H$_\alpha$ in absorption is usually not seen in  
galactic C(N) stars. The reason of this is twofold: i) they are cool
(T$_{\rm eff}\leq 3000$ K), so that the H$_\alpha$ feature is very weak, and ii)
the presence of strong blending in the $\lambda 6560$ {\AA} region due to intense molecular lines (because of
their near solar metallicity).
In the case of D461, despite its considerable carbon enhancement, the intensities of
molecular lines 
(mainly CN and C$_2$) are significantly lower in that region of the spectrum.
Indeed, there are carbon stars of CH type in the Galaxy showing H$_\alpha$ in absorption,
but they are metal poor and relatively hot, with effective temperatures
typically larger than $\sim 4000$ K 
(Vanture 1992). A preliminary analysis of the C1 and C3 spectra, both
belonging to Draco and observed in the same run as D461, shows also evident H$_\alpha$
features. In particular, a huge emission line is found in the C1 spectrum,
which is a well known symbiotic CH
star (Munari 1991), while an absorption line is seen in the C3 spectrum, which 
is classified as a J-type carbon star. We do not detect the Li line in these two stars.

The Li abundance has been derived by means of the spectral synthesis method.
LTE has been assumed. A complete list of atomic and molecular lines (CN, C$_2$, CH)
in a $\sim 50$ {\AA} region around the Li doublet was included in the synthesis
(see Abia et al. 2002 for details on this list). Theoretical spectra
were convolved with Gaussian functions with a FWMH $\sim 1$ {\AA} to simulate
the instrumental profile. Figure 1 shows
fits to the spectral region of the lithium doublet for different values of the
Li abundance. With the atmosphere parameters mentioned above, our best
fit to the Li line is obtained with log $\epsilon \rm{(Li)}=3.5$. N-LTE
corrections would increase this value by 0.1-0.2 dex (cf. Abia et al. 1999). 
The main parameter afflicting the Li abundance derived is T$_{\rm{eff}}$. A
variation of $\pm 200$ K implies a change in the Li abundance by $\pm 0.30$
dex.
Uncertainties in the C/O ratio and the metallicity adopted have much less
impact on the estimated Li abundance. For instance, for the same
T$_{\rm{eff}}$ and a metallicity [Fe/H]= $-1.7$, the estimated abundance 
would be log $\epsilon$(Li)$=3.4$. Finally, a variation of $\pm 0.2$ dex in the log
(C/O) have a negligible effect on the estimated Li abundance.
Adding quadratically all the errors, including the uncertainty
in the continuum location, we estimate a total uncertainty in the Li abundance
of $\pm0.4$ dex 
\footnote{systematic errors, as those due to possible N-LTE effects, are not included.}. 
  
In addition to the Li, a search for other chemical
peculiarities, which might prove the nature of D461, has been done. 
The $^{12}$C/$^{13}$C ratio is an important test for evolved stars. A low 
ratio is expected if the envelope material has been exposed to the CNO burning. 
 C(N) stars in the Galaxy typically present carbon isotopic ratios 
in the range 40-60 (Lambert et
al. 1986, but see also Ohnaka \& Tsuji 1996, and Abia et al. 2002). 
$^{13}$C enhancements can be easily detected by comparing the ratios of
$^{13}$C$^{12}$C $\lambda 4744$ {\AA}
to   $^{12}$C$^{12}$C  $\lambda 4737$  {\AA}   in  the blue or the ratios  of
$^{12}$C$^{12}$C $\lambda 6191$ {\AA} to  
$^{13}$C$^{12}$C  $\lambda 6168$ {\AA}, $^{13}$C$^{12}$C $\lambda 6102$ {\AA}
to $^{12}$C$^{12}$C $\lambda  6122$ {\AA}  and  
$^{13}$C$^{14}$N $\lambda  6260$ {\AA}  to $^{12}$C$^{14}$N $\lambda 6206$
{\AA} at larger wavelengths. Unfortunately, our spectrum did not cover these spectral
regions. Then, we tried to derive the carbon isotopic ratio from $^{12,13}$CN
lines at $\sim\lambda 8000$ {\AA} using the same technique as in Abia \& Isern (1996). 
However, because of the low spectral
resolution, only a lower limit has been obtained, namely  $^{12}$C/$^{13}$C$> 40$. This
limit is not much affected by uncertainties in the stellar metallicity 
and/or the C/O ratio, but it is sensitive to the uncertainty in  T$_{\rm{eff}}$: the
lower the temperature the higher the carbon isotopic ratio.
The quoted lower limit is very conservative, because no CN feature
sensitive to the $^{13}$C abundance changes in the $\lambda 8000$ {\AA} region 
can be fitted with a lower $^{12}$C/$^{13}$C ratio. Note that
there are various $^{13}$CN features in the spectrum of D461 that can be adjusted
with a considerably larger carbon isotopic ratio (i.e. $\sim 100$). This latter value 
would be in better agreement with the theoretical predictions for the 
$^{12}$C/$^{13}$C ratio in this star (see next section).

We have also searched for $s$-element enhancements, a
clear signature of the AGB nucleosynthesis. 
The best spectral region for such analysis ($4750-4950$ {\AA}) is not
accessible in our spectra either and we could only use the Ba II line at 
$\lambda 6497$ {\AA}. This line is strong and usually overcrowded with molecular
lines in carbon stars of near solar metallicity, so that
that it is not useful for abundance analysis in C-stars with such metallicity. However, 
the low metallicity of D461 helps in reducing the blending, but 
the incompleteness of the list of molecular lines used in this
spectral region reduces our capability of reproducing the Ba blend. An upper
limit has been derived: [Ba/Fe]$<+2.0$. This limit decreases by increasing the 
metallicity and the C/O ratio or by reducing T$_{\rm{eff}}$. 
For instance, adopting a metallicity [Fe/H]$=-1.7$ (keeping constant
T$_{\rm{eff}}$ and C/O) the upper limit would be [Ba/Fe]$<+1.8$. It is also possible
to put a lower limit to the Ba abundance. We have assumed that the spectral feature
lacking in the Ba blend was due to a $^{12}$CN line. Then, we artificially vary its
spectroscopic parameters (excitation energy and $gf$-value) until a fit was
obtained to the Ba blend. The resulting lower limit was [Ba/Fe]$>+0.5$ dex.  

To summarize, although the chemical characteristics of D461 have to be
confirmed by a more detailed
analysis based on higher resolution spectra, we can say that this star is certainly
Li-rich, with a  large carbon
enhancement and with a weak evidence of s-process enrichment.

\section{The nature of Draco 461}

Our spectral analysis of D461 reveals several chemical peculiarities worth 
to be discussed and understood. As far as we know, the estimated C/O ratio 
is the largest ever found for a carbon star. Our tests show that 
for any reasonable choice of [Fe/H] 
and T$_{\rm{eff}}$, the C/O ratio in D461 must be larger than 3. This value is  
certainly larger than those typically found in Galactic C-stars, which typically show C/O
only slightly above 1 (Lambert et al. 1986; Abia et al. 2002). 
Recently, Matsuura et al. (2002), by analyzing the intensities of C$_2$H$_2$ and HCN bands,
found somewhat larger C/O ratios ($>1.2$) in some C-stars of the LMC. 

\begin{figure}
\centering
\includegraphics[width=\textwidth]{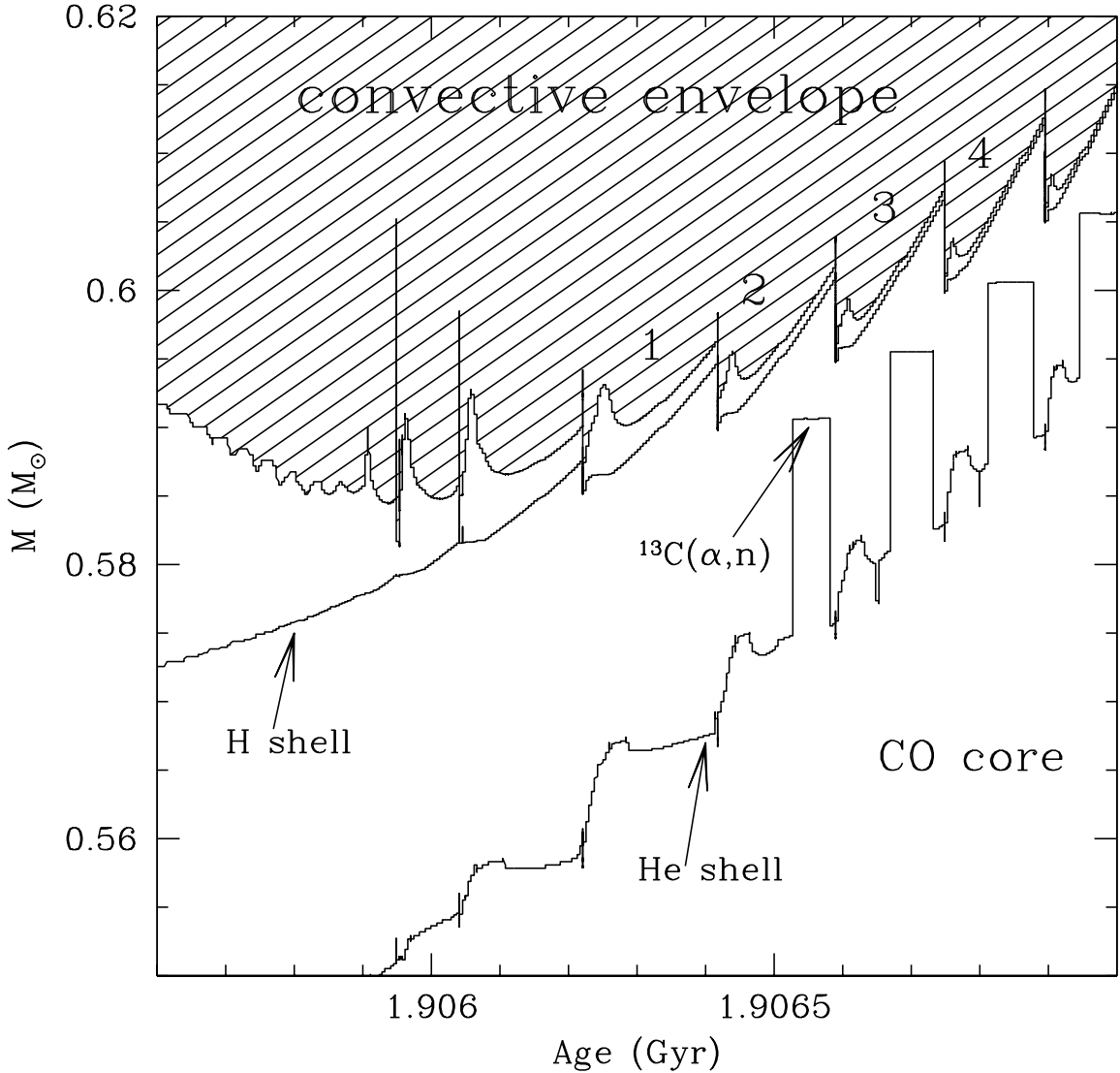}
\caption{Evolution of the locations (in mass coordinates) of the maximum
energy production in the H and He burning shells (solid lines), for a thermally pulsing
AGB stellar model with M=1.5 M$_\odot$ and $Z=0.0003$. 
The hatched area shows the lower portion of the convective envelope.
Interpulse periods, after the first TDU episode, are numbered as in Table 1.
Note that during the interpulse, when the $3\alpha$ is off, the maximum
nuclear energy production moves to the location of the $^{13}$C pocket, where the 
$^{13}$C$(\alpha,n)^{16}$O reaction and the related s-process nucleosynthesis take place.}   
\label{}%
\end{figure}

   \begin{table}
    \caption{Theoretical predictions for the surface composition of D461}
     \label{label}
    $$
      \begin{array}{lccccccccc}
            \hline
                  &  & \rm{D461}^{a} & & \rm{Initial}  &  \rm{RGB} &  1st^{b} & 2nd & 3rd & 4th \\
            \hline
            & & & & & & &  & & \\
            \Delta \rm{M_{up}^{c}} & & & &    &    & 3\cdot10^{-4} & 3\cdot10^{-3} &
4\cdot10^{-3} & 5\cdot10^{-3} \\  
            \hline
            & & & & & & & & & \\
            \rm{log}~\epsilon \rm{(Li)} & &  3.5\pm0.4  & &    &   &  &   &   &   \\
            \hline 
                    std^{d} & & &  & 2.3   & -0.73   & -0.54   & 0.01   & 0.34  & 
0.55  \\
                    20 &   &  &   &  &       & 2.17   &  2.24  &  2.06 &  1.95  \\
                    25 &   &  &   &  &       & 3.29   &  3.39  &   &    \\
                    30 &   &  &   &  &       & 3.62   &  3.76  &   &    \\
                    40 &   &  &   &  &       & 3.92   &  3.76  & 3.68  &    \\
             \hline
            & & & & & & & & & \\
         ^{12}{\rm C}/^{16}{\rm O} &  & 3-5  &  &   &  & &    &   &   \\
             \hline
                    std & & & & 0.50  & 0.32   & 2.10   & 10.44   & 18.06  & 23.13  
\\
                    20  & &  & &    &        & 2.10   & 10.44   & 18.06  & 23.13  
\\
                    25  & &  & &    &        & 2.10   & 10.44   &   &    \\
                    30  & &  & &    &        & 2.09   & 10.44   &   &    \\
                    40  & &  & &    &        & 1.79   &  9.62   & 15.79  &    \\
                 \hline
            & & & & & & & & & \\
        ^{12}{\rm C}/^{13}{\rm C} &   & > 40 &  &      &    &   &    &   &   \\
               \hline
                    std & & & & 89.9  &  22.3  & 146.7   &  895.6  & 2053.7   &
3394.7   \\
                    20  &  &  & &   &        & 146.7   &  880.7  & 1990.4  & 
3264.5  \\
                    25  &  &   & &  &        & 141.7   &  768.3  &   &    \\
                    30  &  &   & &  &        &  98.3   &  238.1  &   &    \\
                    40  &  &   & &  &        &  18.2   &   40.2  & 27.6  &    \\
              \hline
            & & & & & & & & & \\
         $[Ba/Fe]$ & & 0.5-2.0 &   &   & &  &    &   \\
           \hline
                   & & & & 0.  & 0. & 10^{-5}  & 1.37  &  1.83 &  2.05  \\
            \hline
           \end{array}
$$
\begin{list}{}{}
\item[{$^{a}$}] Estimated values for D461

\item[{$^{b}$}] Interpulse numbers corresponding to the labels reported in Figure
2. 

\item[{$^{c}$}] Mass of dredged up material (in M$_\odot$).

\item[{$^{d}$}] $std$ refers to models without CBP (see text); 20, 25, 30 and 40
indicate T$_{\rm{CBP}}$, the maximum temperature of the CBP, in 10$^6$ K.

\end{list}
\end{table}

\begin{figure}
\centering
\includegraphics[width=\textwidth]{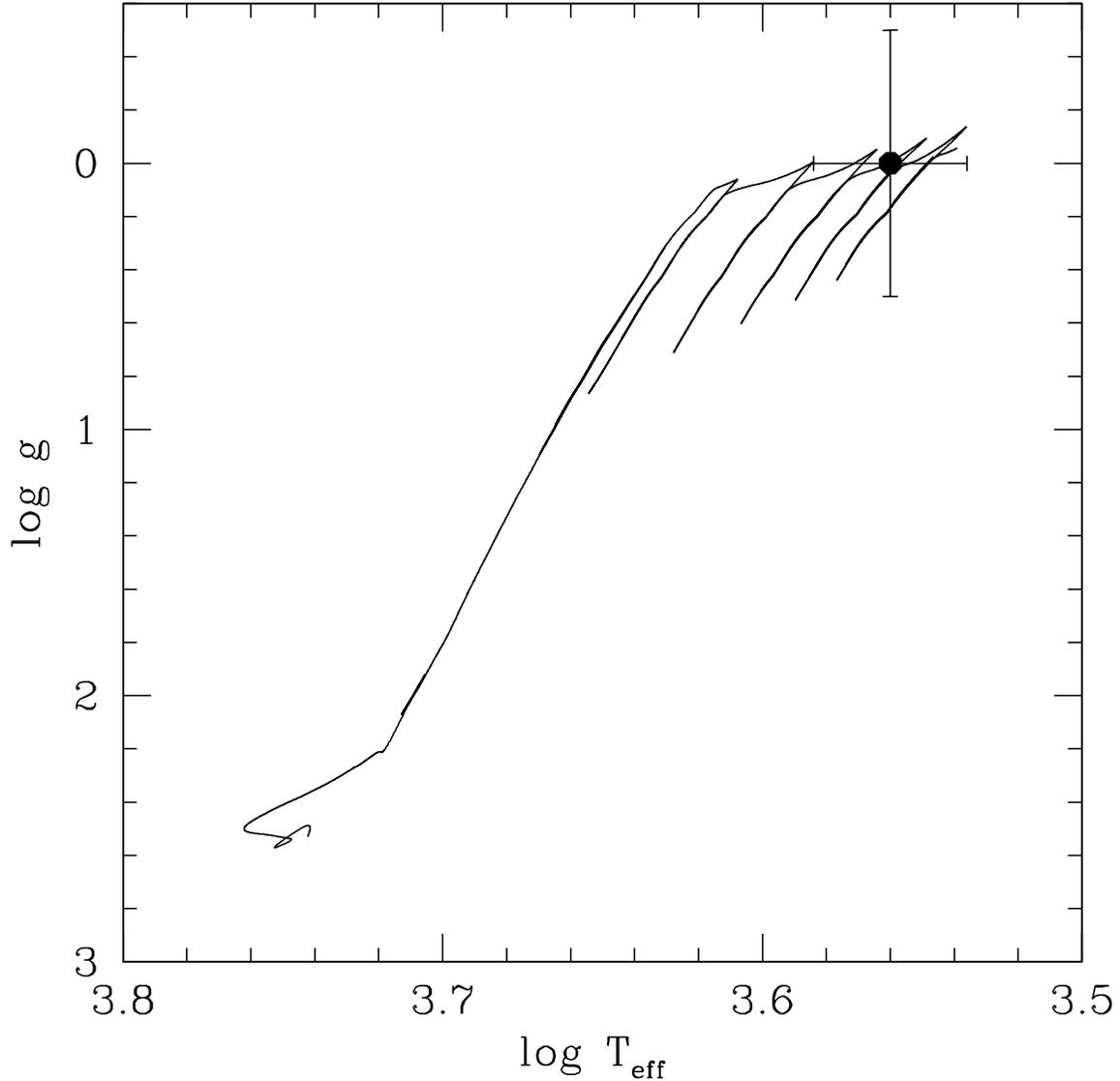}
\caption{Evolutionary track of 1.5 M$_\odot$ (Z=0.0003) stellar model in the
log $g$-log T$_{\rm{eff}}$ diagram. The track starts at the beginning of the core He burning. 
Few thermal pulses are included. The position of D461 is also shown. It is consistent with
a thermally pulsing low mass AGB undergoing the third dredge up.} 
\label{}%
\end{figure}

The C enhancement may be due to the  
third dredge up (TDU) occurring in thermally pulsing AGB stars.
As it is well known, the TDU brings to the surface
the ashes of the He burning (essentially carbon) and the products of the neutron 
capture nucleosynthesis (s-process). 
 The detection of Tc, whose s-component has lifetime  
$\sim 2\cdot10^5$ yr, in galactic C(N) stars proves that
they are actually undergoing the third dredge up.
Nevertheless, some C-stars could not be AGB stars experiencing the TDU.
This may happen in interacting binary systems, when the
atmosphere of the secondary (less evolved) star is polluted with the matter
lost by an AGB companion (Vanture 1992; Jorissen 1999).
 The accretion of the less massive component may occur 
as a consequence of the Roche lobe overflow of the primary (in close systems)
or by wind (in loose systems). C-stars of this type are called {\it extrinsic},
to be distinguished from the {\it intrinsic} C-stars that are AGB stars undergoing the TDU.
 The depth of the TDU depends on the envelope mass (Iben \& Renzini 1983).
Current theoretical models show that there exists a minimum initial mass 
for the occurrence of the TDU, which varies
between 1.3 and 1.5 M$_\odot$, depending on 
the metallicity and on the pre-AGB mass loss rate (see e.g. Straniero et al. 2003). 
This is confirmed by the observational evidence that intrinsic C-stars are not 
found in very old stellar systems, where 
stars with mass exceeding this limit are already evolved beyond the AGB. 
The galactic halo is an example of these old stellar populations, 
whose carbon rich stars (CH and dwarf-C) are expected to be extrinsic (Wallerstein \& Knapp 1998). 
At the metallicity of Draco, even considering a 
particularly moderate RGB mass loss, the minimum mass for the occurrence of the TDU 
cannot be lower than 1.2 M$_\odot$, which corresponds to an AGB age $<4$ Gyr.   
As in the case of the halo, however, the dominant stellar population in Draco 
is very old, namely $\sim10$ Gyr (Grillmair et al. 1998).  
Nevertheless, the abundance pattern of the Draco stellar population differs from
that of halo stars with similar metallicity, the most striking peculiarity 
being the nearly solar, in some cases only slightly enhanced [$\alpha$/Fe] ratios 
(Shetrone et al. 1998, 2001a and 2001b). In addition, 
the presence of anomalous Cepheids in Draco (Baade \& Swope 1961; Gallart et al. 1999)
demonstrates the existence of a  more massive (1-2 M$_\odot$) 
and younger stellar population. 
It appears, therefore, that the star formation in Draco would have been extended 
up to a few Gyrs ago (Aparicio et al. 2001; Ikuta \& Arimoto 2002).
Thus, in spite of the large age of its dominant stellar population,
 it is possible that some C-stars of Draco could be intrinsic AGB stars with masses 
 of the order of 1.5 M$_\odot$. 

By comparing the available photometric and spectroscopic data to 
extant stellar models of AGB stars, we have looked into the hypothesis of
the intrinsic nature of the C enhancement of D461. 
Then, an evolutionary sequence of models for a star with
M$=1.5$ M$_\odot$, $Z=3\cdot10^{-4}$ and Y$=0.24$ have been computed using the FRANEC code 
(Chieffi et al. 1998). 
The calculation was started from the pre main sequence up to the AGB phase. 
A Reimer's mass loss rate ($\eta=0.4$) has been
adopted since the RGB phase. A time-dependent mixing in the convective zones have been obtained
following the algorithm described by Chieffi et al. (2001). At the convective boundaries, 
an exponential decay of the convective velocity is assumed, which allows the formation 
of a $^{13}$C pocket at the epoch of the TDU (see Cristallo et al. 2004 for further details).   
Few thermal
pulses have been followed in details. 
In particular, in order to evaluate
the nucleosynthesis taking places in the H and He burnings shells (s-process included),
a full network of 450 isotopes (about 700 reactions) has been coupled to the stellar 
structure equations. 
Figure 2 shows the evolution  of the locations (in mass coordinates) of the H and He burning
shells during the TP-AGB phase. 
The position of the bottom of the convective envelope is also reported. 
The dredge up occurs when the inflation powered by a thermal pulse (TP) cools the base of
the envelope and the H burning dies down.
Starting from
the second TP, the bottom of the convective envelope regularly penetrates
into the H-exhausted core, down to the region previously enriched with the
products of the He burning \footnote{The thermal pulse generates a convective zone that extends
from the He burning shell up to the top of the H-exhausted region. Thus, C and s-process elements are 
efficiently mixed within this convective zone.}. The amount of mass dredged up is reported in Table 1.
After a dredge up episode, the  expanded layers fall back, the H burning restarts and 
the convective envelope recedes. Close to the zone of 
maximum penetration, the receding envelope leaves a variable profile of H. At the H re-ignition,
a $^{13}$C pocket forms in this zone.  The total mass of $^{13}$C within the pocket ranges
between 10$^{-5}$ M$_\odot$ (after the first TDU episode) and 10$^{-6}$ M$_\odot$ 
(last computed TDU).
Then, during the interpulse, the temperature rises up, until
the  $^{13}$C$(\alpha,n)^{16}$O reaction takes place. This neutron source activates the
s-process nucleosynthesis and then, heavy isotopes (up to Bi) are synthesized (Straniero et al. 1995;
Gallino et al. 1998; Cristallo et al. 2004).  
The resulting modification of the envelope composition 
is reported in Table 1. 
 
The evolutionary track in the log g$-$log T$_{\rm{eff}}$ diagram is shown in
Figure 3. The position of D461, as derived by best fitting the
spectra, is also reported.
The effective temperature of the models drops after the onset of the third dredge up.
This is due to the increase of the envelope opacity caused by the carbon enhancement 
(Marigo et al. 2002) \footnote{We have
simulated the change of the opacity caused by the dredge up by means of opacity tables with 
variable metallicity, but scaled solar mixture. This probably underestimates the 
contribution of some molecular
species that form in a C rich mixture. Unfortunately, the available low temperature
opacity tables does not include such effects.}. The figure shows that
the effective temperature of the AGB model approaches that estimated for  D461
after a few dredge up
episodes. The estimated gravity is also compatible with that of  low mass TP-AGB stellar models.
Note that due to the scarcity of O in the envelope, the C/O ratio immediately 
grows above the unity
(see Table 1). After two dredge up episodes the C/O is about 10.
 Here we have assumed scaled solar initial composition.
An oxygen enhancement of a factor 2 or 3, or a larger metallicity of the initial model,
would lead to a better agreement with the estimate of the C/O ratio 
for D461. The lower limit for the carbon isotopic ratio and the constraint for the 
[Ba/Fe] (see the previous section) are also fulfilled.
On the contrary the lithium production cannot be explained with our standard model.
This is because the temperature at the base of the convective envelope never attains
 $2\cdot10^7$ K, the minimum temperature for the synthesis of Be.

This is a well known problem: some peculiar low mass AGB stars show low $^{12}$C/$^{13}$C
and high Li abundances, which are clear signatures
of nuclear processes occurring at the base of the convective envelope (Abia \& Isern 1997).  
To solve such kind of problems, Wasserburg et al. (1995; see also Nollet et al. 2003) suggested
 that some amount of material could be transported
from the fully convective envelope into the underlying radiative
region, down to the outer zone of the H burning shell.
Although the actual physical basis for such phenomenon
 (that they called {\it cool bottom process}, CBP)
 is not known, it has been invoked to explain some chemical anomalies
of RGB stars
and in studies of dust grains that formed in circumstellar
envelopes (e.g., Huss et al. 1994; Boothroyd et al. 1994; Harris et al. 1985; Kahane
et al. 1992). 

Then we have computed some additional AGB stellar models by including a parameterized
description of the deep circulation during the interpluse period. 
Following the 
scheme proposed by Nollet et al. (2003),
we assume that the CBP depends on two free
parameters, namely the maximum temperature reached by the 
circulating material (T$_{\rm{CBP}}$) and the circulation velocity (v$_{\rm{CBP}}$) \footnote{
Nollet et al. actually use the mass flux instead of the circulation velocity, 
but these two quantities are clearly related.
Note that in Nollet et al. the effect of the CBP on the nucleosynthesis 
was evaluated by means of a post-process, whereas we have included this calculation 
in the stellar evolution code. In particular, we mimic the CBP with a deep-mixing obtained
by means of the same algorithm used to describe the mixing in 
the convective zones (see Chieffi et al. 2001 for details). Thus, the extension
and the efficiency of the deep-mixing are controlled by T$_{\rm{CBP}}$ and v$_{\rm{CBP}}$,
respectively.  
In agreement with the previous investigation, we find 
that the key parameter is the maximum temperature reached by the 
circulating material.  
T$_{\rm{CBP}}$ has been varied from 20 to 40 ($10^6$ K). 
Within this range the maximum penetration in mass is of the order of
5 $10^{-3}$ M$_\odot$ and the physical structure 
of the star is very marginally affected. 
 In the zone where the CBP takes place, we have adopted a constant circulation velocity 
equal to a fraction (1, 1/10, 1/100) of the mixing
velocity in the most internal layers of the convective envelope. 
Typically the maximum explored values for  v$_{\rm{CBP}}$ are of the order of 4 $10^{4}$ cm/s, 
which correspond to a mass flux of 0.05  M$_\odot$/yr. The minimum value is about 
100 times smaller. 
Within this range, the resulting surface Li abundance is practically insensitive to a change of 
v$_{\rm{CBP}}$.}
       
The effect on the surface composition 
is illustrated in Table 1. Models including the CBP are identified by the corresponding 
T$_{\rm{CBP}}$. The standard ($std$) model refers to the case without CBP. As a consequence of the 
cool bottom process,
the abundance of Li initially increases, reaches a maximum and then,
once the $^3$He (the fuel for the Be production) is consumed, decreases.
The better agreement with the measured Li abundance in D461 is obtained for a bottom temperature 
of 25-30 ($10^6$ K). Another consequence of the CBP is the partial consumption of $^{12}$C
and the enhancement of the $^{13}$C in the envelope.
 This affects the carbon isotopic ratio and the C/O, which remain, in any case, within the
 derived observational constraints for D461.

\section{Conclusions}

Our analysis has been limited by the low resolution of the spectra and by the lack of
reliable model atmospheres appropriate for the peculiar chemical composition of D461.
The C/O, for example, is significantly larger than that measured in any other C-star. 
In spite of that, several features in the spectra of D461 converge toward the same
conclusion: it is a thermally pulsing
low mass ($\sim 1.5$ M$_\odot$) low metallicity ([Fe/H]$\sim$ $\rm{-2)}$ AGB star, undergoing
the third dredge up.
Gravity and effective temperature are also consistent with this hypothesis.
This is the first discovery of an intrinsic C(N) star in a 
metal poor stellar population.    
The Li enhancement is likely due to a moderate cool bottom process.

This scenario implies that the mass of D461 cannot be smaller than 1.3 M$_\odot$, because 
a less massive star would attain the thermally pulsing AGB phase with a
too small envelope for the occurrence of the third dredge up.
On the other hand, the observed visual magnitude (nearly that of the RGB tip) 
excludes the possibility that D461 could be a massive AGB star.
We estimate that 2 M$_\odot$ is a conservative upper limit for the mass of D461.
Such a stringent limitation of the stellar mass implies a striking constraint to
the age of D461: it should range between 1 and 3 Gyr 
(see e.g.  Table 1 in Dom\'\i nguez et al. 1999). 

We remind, again, that the bulk of the stellar population of the
Draco dwarf spheroidal is dominated by very old stars 
($\sim 10$ Gyr) although the star formation extended up to 2 Gyr ago.
This latter corresponds to the age of a 1.5 M$_\odot$  AGB star (for the typical metallicities
of Draco).
Note that the anomalous Cepheids, 
which are currently detected in Draco, 
should have masses very close to this value (Bono et al. 1997). 

Let us finally mention an alternative scenario. It could be possible that D461
was, in origin, a low mass star (M$<$1 M$_\odot$) 
whose mass has been recently
accreted,  as a consequence of a mass exchange or coalescence, 
from a companion in a close binary system. Up to
about 200 Myr ago D461 was an old main sequence star becoming
a Blue Straggler and, more recently, a pulsating horizontal branch star
with the characteristics of the anomalous Cepheids.
Radial velocity monitoring of D461, by Olszewski et al. (1996), found no evidence 
of the existence of a compact companion, but
Shetrone et al. (2001a) detected a possible photometric variability 
(at 4 $\sigma$ level). Such a variability is, however, quite common among
AGB stars belonging to intermediate-age populations. Additionally, note that 
 the binary hypothesis is disfavoured because of the high Li abundance (see Abia et al. 1993). 
In summary, the presently available data cannot allow a definitive discrimination between 
the old and the young hypothesis for D461, even if, from a statistical point of view,  
the latter is the more significative.

\begin{acknowledgements}
Data from the VALD  database at Vienna were used for the preparation of this paper. 
The 4.2m WHT is operated on the island of La  Palma by the  RGO in the
Spanish Observatory of the Roque de los Muchachos of the
Instituto de Astrof\'\i sica de Canarias. This work has been partially  
supported by the MCyT grant AYA2002-04094-C03-03 and by the spanish-italian 
cooperation INFN-CICYT.

\end{acknowledgements}

\end{document}